\documentclass{article}
\usepackage{amsmath}
\usepackage{graphicx}
\usepackage[hidelinks]{hyperref}
\usepackage{float}
\usepackage{threeparttable}
\usepackage{natbib}
\usepackage{url} 
\usepackage{graphicx}
\usepackage{amsthm,amsmath,amsfonts,amssymb}
\usepackage{mathtools}
\usepackage{amscd,amssymb,verbatim}
\usepackage{mathrsfs}
\usepackage{bbm}
\usepackage{listings}
\usepackage{epsfig}
\usepackage{enumitem}
\usepackage[plain]{algorithm2e}
\usepackage{algorithmic}
\usepackage{url}
\usepackage{booktabs,makecell,ltablex}
\usepackage{multirow}
\usepackage{moreverb}
\usepackage{xcolor}
\usepackage{setspace}
\usepackage{subcaption} 
\usepackage[all,cmtip]{xy}
\usepackage{rotating}

\graphicspath{{Figs/}}
\makeatother

\usepackage{tikz}
\usetikzlibrary{positioning,arrows.meta}

\begin{document}

\def\spacingset#1{\renewcommand{\baselinestretch}%
{#1}\small\normalsize}
\singlespacing


\title{\bf Bayesian hierarchical bootstrap framework for causal subgroup estimation with a time-to-event outcome}
\author{Mengyao Shi\thanks{Dalla Lana School of Public Health, University of Toronto, Ontario, Canada.}, \hspace{.2cm}
Amanda Ricciuto\thanks{Division of Gastroenterology, Hepatology Nutrition, The Hospital for Sick Children, Ontario, Canada.}, \hspace{.2cm} Mark Deneau\thanks{University of Utah and Intermountain
Healthcare Primary Children's Hospital, Utah, United States.}, \hspace{.2cm} Kuan Liu\thanks{Institute of Health Policy, Management and Evaluation, University of Toronto, Ontario, Canada.} \hspace{.2cm} \\
  }
 \date{ }
  \maketitle

\bigskip
\begin{abstract}
{Causal estimation of treatment effects within prespecified subgroups, such as biomarker-defined strata, disease phenotypes, or demographic groups are often of clinical interest.  Bayesian approaches are attractive for subgroup effect estimation because flexible priors can represent complex treatment heterogeneity and propagate posterior uncertainty. Frequentist methods for prespecified causal subgroup analysis with time-to-event outcomes are also available, including propensity-score weighting approaches that emphasize subgroup-level covariate balance \cite{yang2023propensity}. In posterior g-formula, subgroup survival estimands also depend on the subgroup specific distribution of baseline covariates, which may be unstable when some subgroups are small or unevenly represented. Right censoring separately reduces information for the outcome-model component of the estimand, increasing overall uncertainty in subgroup causal survival contrasts. We extend the hierarchical Bayesian bootstrap (HBB) to subgroup causal inference with right-censored time-to-event outcomes. The HBB places a nonparametric hierarchical prior on subgroup specific baseline covariate distributions, enabling principled borrowing of information across related subgroups while preserving subgroup specific structure. We combine this distributional regularization with a Bayesian accelerated failure time model and right censoring to perform posterior g-formula that propagates uncertainty from both the survival model and the subgroup covariate distribution. The resulting framework stabilizes subgroup causal survival estimands in sparse strata without imposing parametric assumptions on the covariate distribution. Simulation studies examine performance across varying degrees of subgroup sparsity and censoring and compare the proposed approach to the popular Bayesian additive regression tree for heterogeneous survival effects.
 }

\end{abstract}

\noindent%
{\it Keywords:} time-to-event data, heterogeneous treatment effect, informative censoring, causal inference, non-parametric Bayesian estimation, Bayesian additive regression trees
\vfill

\newpage
\spacingset{1.5}

\section{Introduction}\label{sec1}
It is increasingly recognized in causal inference that treatment effects may vary across individuals and prespecified subgroups, such that population-level effects can coexist with heterogeneous subgroup-specific effects. In clinical studies, there is often interest in estimating treatment effects within prespecified subgroups defined by characteristics such as sex, age group, or baseline disease status. Under standard identification assumptions, these subgroup effects are typically defined through standardization, which highlights that reliable inference requires not only flexible outcome modeling but also stable estimation of the subgroup specific covariate distributions. The g-formula \cite{ROBINS1986,ROBINS1987139S,ROBINS1987} provides a formal framework for identifying subgroup causal estimands in time-to-event settings by expressing counterfactual survival quantities as expectations over the subgroup-specific distribution of baseline covariates. This representation makes explicit that valid subgroup causal inference involves two distinct components: the estimation of the conditional survival response function and the characterization of the covariate distribution within each subgroup.

Recent Bayesian causal inference methods have enabled flexible modeling of conditional outcome distributions for estimating marginal and heterogeneous treatment effects. Specifically, Bayesian additive regression trees (BART) were used in early work to estimate marginal treatment effects \cite{Hill2011} and were subsequently extended to estimate individual and conditional average treatment effects \cite{zeldow2019,hahn2020}. Closely related approaches based on Bayesian causal forests further adapted tree-based priors to treatment effect learning by incorporating regularization informed by treatment assignment mechanisms \cite{caron2022,starling2021targeted}. Beyond tree-based priors, a broad class of Bayesian nonparametric models, including Dirichlet process mixtures and their extensions, has been developed for posterior inference on causal estimands in complex data structures and causal settings, including zero-inflated outcomes \cite{oganisian2020bayesian}, missing data \cite{roy2018bayesian}, mediation analysis \cite{kim2016framework}, survival outcomes with competing risks \cite{xu2020bayesian}, and causal quantile effects \cite{xu2018bayesian}.

Bayesian causal methods developed for time-to-event outcomes are particularly relevant to the present work. Flexible Bayesian survival models, including accelerated failure time (AFT) formulations based on causal BART, have been developed to estimate heterogeneous treatment effects on survival outcomes under relatively weak parametric assumptions, while accommodating complex covariate effects, interactions, and departures from proportional hazards \cite{Henderson2020-vb,Hu_2021,Hu2020EstimatingHS}. This line of work demonstrates the potential of flexible Bayesian survival models to estimate complex outcome surfaces and variation in treatment response.

However, the primary source of regularization in these approaches is the conditional survival response surface. Subgroup causal survival estimands are obtained through g-computation by averaging model-based counterfactual survival predictions over subgroup-specific covariate distributions, which are commonly represented using empirical plug-in distributions \cite{Henderson2020-vb}. When the empirical subgroup covariate distribution is treated as fixed, posterior uncertainty in that distribution is not explicitly represented or regularized. This limitation may be particularly consequential in survival settings because censoring reduces the information available for estimating the conditional survival model, which can compound instability in subgroup contrasts when subgroup sample sizes are small or covariate support is limited. Thus, in implementations that condition on fixed empirical subgroup covariate distributions, uncertainty in the outcome model is represented, whereas uncertainty in the subgroup covariate distribution remains unmodeled.

Recent work has proposed hierarchical Bayesian frameworks that address this challenge by borrowing information across related subgroups, with the aim of improving stability when empirical subgroup distributions are estimated from limited data \cite{Arman2024}. Specifically, the hierarchical Bayesian bootstrap (HBB) treats subgroup covariate distributions as inferential quantities that can share information across subgroups, rather than as fixed empirical distributions estimated independently within each subgroup, thereby providing a means of stabilizing subgroup-specific causal effect estimation under the g-formula \cite{Arman2024}.

Because the g-formula expresses subgroup causal estimands as expectations over subgroup-specific covariate distributions \cite{Chen_2025}, it is natural to address instability by introducing structure at the distributional level. Hierarchical Bayesian modeling enables partial pooling across subgroups while preserving subgroup-specific characteristics. The HBB applies this principle by extending Rubin's Bayesian bootstrap to stratified settings, with subgroup-specific distributions partially pooled toward a shared population-level distribution. Subgroups with substantial support behave similarly to standard Bayesian bootstrap estimators, whereas small or sparsely supported subgroups receive stronger borrowing. This partial pooling can reduce variability in sparsely represented subgroups without imposing a parametric model for the covariate distribution. From the perspective of subgroup causal inference, HBB directly targets the distributional component required by the g-formula. When subgroup causal estimands are expressed as averages of counterfactual survival responses over subgroup distributions, uncertainty in those distributions contributes materially to uncertainty in the estimand itself. HBB provides a Bayesian representation of uncertainty in the subgroup covariate distributions and propagates this uncertainty through g-formula. In settings with small or imbalanced strata, this additional structure may reduce the instability that arises when subgroup distributions are represented using fixed empirical plug-in distributions. These considerations are also relevant in time-to-event analysis. Censoring reduces the information available for estimating the conditional survival distribution and therefore increases uncertainty in subgroup survival contrasts. Extending HBB ideas to survival outcomes therefore offers a framework that may improve the stability of subgroup causal survival inference relative to approaches that regularize only the outcome model, particularly when subgroup covariate distributions are estimated from limited data.

Despite these developments, HBB has not yet been extended to time-to-event causal inference, where censoring introduces an additional source of uncertainty. Many existing Bayesian survival methods provide flexible outcome modeling while conditioning on empirical subgroup covariate distributions, whereas existing HBB frameworks regularize subgroup distributions but have been developed primarily for non-survival outcomes. To our knowledge, no existing framework jointly combines flexible Bayesian survival modeling with hierarchical regularization of subgroup covariate distributions and propagation of uncertainty from both components of the g-formula. The present work advances this line of research by extending HBB to subgroup causal effect estimation with survival data under an AFT framework. The proposed framework accommodates censoring in a principled manner, ensuring that the reduction in effective information is reflected in both posterior uncertainty and borrowing behavior. Our work provides a unified approach to subgroup causal survival inference that integrates hierarchical distributional borrowing with flexible survival modeling in a way that has not been addressed by existing methods.

\section{Causal framework for subgroup effect estimation}

\subsection{Causal estimand and assumptions}
\label{sec:causal_estimand}

For subject \(i=1,\ldots,n\), \(A_i\in\{0,1\}\) denotes the treatment received at baseline, \(X_i\) the measured baseline
covariates, and \(G_i\in\{1,\ldots,K\}\) membership in one of \(K\) prespecified subgroups. Let \(T_i^*(a)\) be the potential event time under treatment \(a\in\{0,1\}\). The event time under the observed treatment is therefore \(T_i^*=T_i^*(A_i)\). With \(C_i\) denoting the censoring time, the observed follow-up consists of \(Y_i=\min\{T_i^*,C_i\}\) and \(\Delta_i=\mathbb{I}(T_i^*\leq C_i)\).

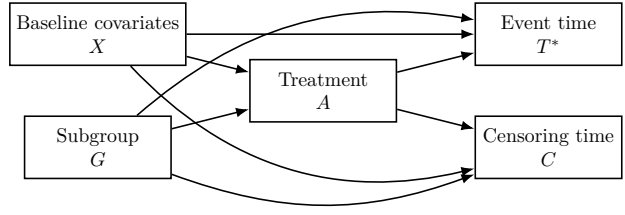
\begin{figure}[t]
\centering
\resizebox{0.48\textwidth}{!}{
\begin{tikzpicture}[
  >=Latex,
  every node/.style={
    draw,
    rectangle,
    minimum width=2.6cm,
    minimum height=1.1cm,
    align=center
  },
  line width=0.8pt
]

\node (X) at (0,2) {Baseline covariates\\\(X\)};
\node (G) at (0,0) {Subgroup\\\(G\)};
\node (A) at (4,1) {Treatment\\\(A\)};
\node (T) at (8,2) {Event time\\\(T^*\)};
\node (C) at (8,0) {Censoring time\\\(C\)};

\draw[->] (X) -- (A);
\draw[->] (G) -- (A);
\draw[->] (X) -- (T);
\draw[->, bend left=25] (G) to (T);
\draw[->] (A) -- (T);

\draw[->] (A) -- (C);
\draw[->, bend right=30] (X) to (C);
\draw[->, bend right=20] (G) to (C);

\end{tikzpicture}
}
\caption{Assumed causal structure for subgroup treatment-effect
estimation with a right-censored event time.}
\label{fig:dag}
\end{figure}

Subgroup membership is a baseline characteristic used to define the populations in which treatment effects are evaluated; it is not itself a target of intervention. The causal structure in
Figure~\ref{fig:dag} allows treatment assignment and the time event distribution to vary across subgroups, as indicated by the paths \(G\rightarrow A\) and \(G\rightarrow T^*\). The censoring process may also depend on treatment and baseline characteristics, while occurring after treatment initiation. For treatment level \(a\) and subgroup \(g\), the target estimand is the potential survival probability at time \(t\),
\begin{equation}
S_g^a(t) = \Pr\!\left\{T_i^*(a)>t\mid G_i=g\right\}.
\label{eq:estimand}
\end{equation}
Comparing \(S_g^1(t)\) with \(S_g^0(t)\) characterizes how the treatment effect varies across prespecified subgroups. Identification of this estimand rests on four assumptions.

\begin{enumerate}[
label=\textbf{Assumption \arabic*.},
leftmargin=*,
widest=4,
align=left
]

\item \textbf{Consistency.}\label{assump:consistency}
For an individual who receives treatment \(a\), the observed event time
agrees with the corresponding potential event time:
\(A_i=a\Rightarrow T_i^*=T_i^*(a)\).

\item \textbf{Conditional exchangeability.}
\label{assump:exchangeability}
Within levels of the measured baseline covariates and subgroup
membership, treatment assignment is independent of the potential event
time; that is,
\(T_i^*(a)\perp\!\!\!\perp A_i\mid X_i,G_i\) for
\(a\in\{0,1\}\).

\item \textbf{Positivity within subgroups.}
\label{assump:positivity}
Each treatment level has a positive probability of being observed for
every covariate profile represented in subgroup \(g\):
\(\Pr(A_i=a\mid X_i=X,G_i=g)>0\) for all \(X\) in the support of
\(X_i\mid G_i=g\).

\item \textbf{Conditionally independent censoring and censoring positivity.}
\label{assump:censoring}
Conditional on treatment, measured baseline covariates, and subgroup membership, the event and censoring times are independent,
\(T_i^*\perp\!\!\!\perp C_i\mid A_i,X_i,G_i\). Follow-up also
remains possible over the time range of interest, so that
\(\Pr(C_i\geq t\mid A_i=a,X_i=X,G_i=g)>0\).

\end{enumerate}

The first three assumptions link the potential survival probability to the observed conditional event time distribution. Standardizing this
distribution over the baseline covariate distribution in subgroup
\(g\) gives
\begin{equation}
S_g^a(t)
=
\int
S_{T^*}
\!\left(
t\mid A=a,X=x,G=g
\right)
\,dF_g(x),
\label{eq:identified_gformula}
\end{equation}
where \(F_g\) is the distribution of \(X\) among individuals with \(G=g\). The fourth assumption permits the conditional survival function in
Equation~\eqref{eq:identified_gformula} to be estimated from the right censored observations where the observed events contribute through the time event density and the censored follow-up contributes through the corresponding survival probability.

\subsection{HBB for Subgroup Causal Survival Estimation}

\label{sec:method}

The potential survival probability in~\eqref{eq:identified_gformula} involves two components: the conditional survival function and the subgroup-specific covariate distribution \(F_g\). Bayesian survival models account for uncertainty in the conditional survival function, whereas posterior g-computation often averages survival predictions over a fixed empirical estimate of \(F_g\). This approximation may be unstable when a subgroup is small or has limited covariate support. We instead place a hierarchical nonparametric prior on \(F_g\), allowing covariate information to be shared across subgroups while preserving subgroup-specific variation. The AFT model supplies the conditional survival predictions, and HBB regularizes the distribution used to average them.

Following the hierarchical construction of ~\cite{Arman2024}, we assign
$
F_g\mid\alpha_g,F_0
\sim
\operatorname{DP}(\alpha_gF_0)
$, $
F_0\mid\gamma,F_*
\sim
\operatorname{DP}(\gamma F_*),
$
where \(F_0\) provides the shared distributional support and \(\alpha_g\) controls the extent to which subgroup \(g\) borrows from it. Here, \(\operatorname{DP}\) denotes a Dirichlet process. Under the Bayesian bootstrap specification obtained by setting \(\gamma=0\),
the posterior baseline distribution is supported on the observed covariate values, and
$
F_g\mid F_0,\alpha_g,X_g
\sim
\operatorname{DP}
\left(
\alpha_gF_0+
\sum_{i\in\mathcal{S}_g}\delta_{X_i}
\right),
$
where \(\mathcal{S}_g=\{i:G_i=g\}\) and \(n_g=|\mathcal{S}_g|\). Writing the baseline distribution as

\begin{align}
\mathbb{E}\!\left\{F_g(\cdot)\mid \pi, \alpha_g, X_g\right\} =
&\mathllap{=}\ \frac{\alpha_g}{\alpha_g+n_g} F_0(\cdot)
+ \frac{1}{\alpha_g+n_g} \sum_{i\in S_g} \delta_{X_i}(\cdot) \\
&\mathllap{=}\ \sum_{i=1}^n
\frac{\alpha_g \pi_i + \mathbb{I}(i \in S_g)}{\alpha_g + n_g}
\, \delta_{X_i}(\cdot).
\label{eq:HBB_posterior_mean_derivation}
\end{align}

The concentration parameter controls how strongly \(F_g\) is pulled toward the shared distribution. Following the hyperparameter guidance of \cite{Arman2024}, we set \(\alpha_g=nM/n_g\), where \(n\) is the total sample size and \(M\) is
a user-specified reference size for estimating the covariate distribution within a subgroup. Under the approximation \(\pi_i\approx 1/n\), the relative posterior weight assigned to an observation from subgroup \(g\), compared with an observation supported only through the shared distribution, is
\begin{equation}
\rho_g
=
\frac{\alpha_g\pi_i+1}{\alpha_g\pi_i}
\approx
1+\frac{n_g}{M}.
\end{equation}
when \(n_g\) is small relative to \(M\), this ratio is close to one, so observations within and outside the target subgroup receive more similar weights and borrowing is stronger. As \(n_g\) increases relative to \(M\), observations from the target subgroup receive greater relative weight and the HBB approaches a subgroup-specific
Bayesian bootstrap.
$
\rho_g
=
\frac{n}{\alpha_g}+1
=
\frac{n_g}{M}+1
$
for an atom observed within subgroup \(g\) compared with one supported only through the shared distribution. Consequently, subgroups with \(n_g\ll M\) borrow more heavily, whereas the HBB approaches a subgroup-specific Bayesian bootstrap as \(n_g\) becomes large relative
to \(M\).

We specify an AFT model for the latent event time. The conditional survival function in Equation~\eqref{eq:identified_gformula} is estimated using the AFT model
\begin{equation}
\log T_i^*
=
m_\theta(A_i,X_i,G_i)+\varepsilon_i,
\label{eq:aft_model}
\end{equation}
where \(m_\theta\) allows the time event distribution and treatment response to vary with the measured covariates and subgroup membership. Under ~\ref{assump:censoring}, the contribution of subject \(i\) to the time event likelihood, after omitting terms involving only the censoring distribution, is proportional to
\begin{equation}
L_i(\theta)
\propto
\left\{f_{T^\ast}(Y_i \mid A_i,X_i,G_i;\theta)\right\}^{\Delta_i}
\left\{S_{T^\ast}(Y_i \mid A_i,X_i,G_i;\theta)\right\}^{1-\Delta_i}.
\label{eq:aft_likelihood}
\end{equation}
Together with the AFT specification, this likelihood provides posterior inference for the conditional time event distribution. This likelihood yields valid inference for the time event model under Assumption~4. For any posterior draw $\theta$, the conditional survival functional admits the explicit form
\begin{equation}\label{eq:AFT_survival_derivation}
\begin{aligned}
S_{T^\ast}(t \mid A=a, X=x, G=g; \theta) =
&\mathllap{=}\ \Pr_\theta(T^\ast > t \mid a,x,g) \\
&\mathllap{=}\ \Pr_\theta\!\big(\varepsilon > \log t - m_\theta(a,x,g)\big) \\
&\mathllap{=}\ 1 - F_{\varepsilon,\theta}\!\big(\log t - m_\theta(a,x,g)\big).
\end{aligned}
\end{equation}
where \(F_{\varepsilon,\theta}\) denotes the distribution function of the AFT error term.

\paragraph{Posterior Computation}

Posterior computation retains the distributional sampling steps of the original HBB and combines them with posterior survival predictions from the right-censored AFT model. The shared and subgroup-specific covariate distributions are sampled directly from their Dirichlet posteriors, after which the AFT draws are used to evaluate the subgroup causal survival estimands.

At each Monte Carlo iteration $m=1,\dots,M$, the following steps are performed.

First, we obtain a posterior draw of the shared baseline covariate distribution. Under the HBB specification with $\gamma=0$, this is achieved by drawing
$
(\pi_1^{(m)},\dots,\pi_n^{(m)}) \sim \mathrm{Dirichlet}(1,\dots,1),
$
and forming the random probability measure
$
F_0^{(m)}(\cdot)
= \sum_{i=1}^n \pi_i^{(m)}\,\delta_{X_i}(\cdot).
$

Second, conditional on \(\boldsymbol{\pi}^{(m)}\), we draw the subgroup-specific weights as
\begin{equation}
\boldsymbol{\omega}_g^{(m)}
\sim
\operatorname{Dirichlet}
\left(
\alpha_g\pi_1^{(m)}+\mathbb{I}(1\in\mathcal{S}_g),
\ldots,
\alpha_g\pi_n^{(m)}+\mathbb{I}(n\in\mathcal{S}_g)
\right),
\label{eq:hbb_weight_draw}
\end{equation}
and form \(F_g^{(m)}=\sum_{i=1}^n
\omega_{gi}^{(m)}\delta_{X_i}\).
These weights are sampled directly and do not require an additional Markov chain.

Third, we draw \(\theta^{(m)}\) from the posterior distribution of the AFT model fitted to the right-censored outcome. Because the time event model and the HBB distribution are specified as separate components, the AFT and Dirichlet draws can be generated independently and paired within each Monte Carlo iteration. The conditional survival probabilities are evaluated using
Equation~\eqref{eq:AFT_survival_derivation}.

Finally, the AFT predictions are averaged using the sampled HBB weights. For subgroup \(g\), treatment level \(a\), and time \(t\), the resulting posterior draw is
\begin{equation}
S_g^{a,(m)}(t)
=
\sum_{i=1}^n
\omega_{gi}^{(m)}
S_{T^*}
\!\left(
t\mid A=a,X_i,G=g;\theta^{(m)}
\right).
\label{eq:posterior_gcomp}
\end{equation}

The corresponding subgroup contrast is
\(\Delta_g^{(m)}(t)
=
S_g^{1,(m)}(t)-S_g^{0,(m)}(t)\).

Repeating these steps for \(m=1,\ldots,M\) produces posterior draws of the subgroup survival probabilities and their contrasts. Posterior means and credible intervals are obtained by summarizing these draws. Variation in \(\theta^{(m)}\) reflects uncertainty in the right-censored AFT model, while variation in
\(\boldsymbol{\omega}_g^{(m)}\) carries uncertainty in the
subgroup-specific covariate distribution.

\section{Simulation study}
\label{sec:simulation}

\subsection{Design}

We conducted a simulation study to examine how the representation of the subgroup covariate distribution affects Bayesian estimation of subgroup causal survival probabilities. The comparison included the empirical plug-in estimator (EMP), the within-subgroup Bayesian bootstrap (BB), the hierarchical Bayesian bootstrap (HBB), and a nonparametric
AFT-BART estimator. EMP, BB, and HBB used the same working AFT outcome model and differed only in the distribution over which the conditional survival probabilities were averaged. The AFT-BART estimator provided a more flexible outcome-model comparison.

Each simulated data set contained \(n=300\) individuals from four prespecified subgroups. To vary the amount of information available within each subgroup, we considered allocation probabilities
\((0.35,0.30,0.20,0.15)\), \((0.25,0.25,0.25,0.25)\), and
\((0.38,0.38,0.12,0.12)\). These settings range from moderate imbalance to equal allocation and a design in which observations are concentrated
in the first two subgroups. They allow the consequences of borrowing to be examined as the amount of subgroup-specific information changes.

The baseline covariates consisted of \(X_i=(L_{1i},L_{2i},L_{3i})\). The first two covariates followed standard normal distributions in all
subgroups. We varied the distribution of \(L_{3i}\) to create two forms of covariate support. In the Gaussian setting, \(L_{3i}\) followed the same standard normal distribution in every subgroup. In the Gamma setting, its distribution varied across subgroups, with shape parameters \((8.0,5.67,3.33,1.0)\). The latter setting produces increasing differences in location and skewness across subgroups, with the most distinct distribution occurring in the fourth subgroup. Together, these settings distinguish borrowing among similar covariate distributions from borrowing when the target subgroup has less population level support.

Treatment was generated from a logistic model conditional on \(X_i\) and \(G_i\), allowing both the covariate composition and treatment prevalence to differ across subgroups. Potential event times were generated from a Weibull AFT model with covariate effects, subgroup intercepts, and moderately varying subgroup treatment effects. An unobserved heterogeneity variable \(U_i\), generated independently of \((A_i,X_i,G_i)\), was also included in the time event model. It did not affect treatment assignment and it induces unmeasured heterogeneity.

The primary setting used conditionally independent right censoring. Censoring times were generated from a Weibull AFT model depending on the observed baseline covariates but not on \(U_i\). Thus, although follow-up varied across covariate profiles, the event and censoring times remained independent after conditioning on the variables included in Assumption~4. We also considered a second setting in which \(U_i\) affected both processes and treatment had a subgroup-specific effect on censoring. The shared dependence on \(U_i\), rather than the treatment dependence of censoring itself, induced residual association between the event and censoring times after conditioning on \((A_i,X_i,G_i)\). Thus, this setting evaluates the effect of a departure from conditional independent censoring and it does not represent a censoring mechanism accommodated by the proposed method. For each data set, we estimated the subgroup-specific survival probability difference \(S_g^1(t^*)-S_g^0(t^*)\) at \(t^*=2\). The true value was calculated from the data-generating time event distribution within each subgroup, including the variation induced by \(U_i\). The R code used to reproduce the simulation study is available at \url{https://github.com/mengyaoshi00-code/Bayesian-hierarchical-bootstrap-framework-for-causal-subgroup-estimation-with-time-to-event-outcomes}.

\subsection{Estimators and comparative performance measures}

The same Bayesian Weibull AFT model was fitted for EMP, BB, and HBB. It included the observed baseline covariates, subgroup-specific intercepts, and subgroup-specific treatment effects. Right-censored observations contributed through the survival probability. Since \(U_i\) was not observed, it was not included in the fitted outcome model. This AFT model used consistently across these three estimators.

Posterior treatment-specific survival probabilities were standardized using the three representations of the subgroup covariate distribution introduced in Section~\ref{sec:method}. EMP assigned equal weight to the observed covariate values in the target subgroup. BB replaced these fixed weights with subgroup-specific Bayesian bootstrap draws but retained support only on observations from that subgroup. HBB allowed the target subgroup distribution to place mass on covariate values
observed elsewhere in the sample through the hierarchical weights. We used the calibration described in Section~\ref{sec:method}, with the reference subgroup size fixed at \(15\), so that smaller subgroups borrowed more strongly from the shared distribution. The AFT-BART estimator was fitted separately within the two treatment groups. For each posterior draw, treatment-specific survival probabilities were predicted under both treatment levels and averaged within the target subgroup. This estimator does not use the HBB representation and was included to compare the proposed approach with a more flexible model for the conditional time event distribution.

Each comparative estimators was evaluated over \(1{,}000\) simulated data sets using bias, root mean squared error, average width of the \(95\%\) credible interval, and empirical coverage. We also report the mean squared error relative to HBB, with values greater than one indicating a lower mean squared error for HBB. The complete data-generating coefficients, prior specifications, computational settings, and results under the alternative censoring mechanism are provided in Supplementary Material~1.


\subsection{Main Results}
\label{sec:results}

\begin{table}[!htbp]
\centering
\caption{Simulation results under non-informative censoring across three subgroup allocation scenarios and two covariate settings. Results are based on $S=1{,}000$ replications with $N=300$ subjects and $K=4$ subgroups. Strata~1 and~4 represent the largest and smallest subgroups under the decreasing and front-heavy scenarios, respectively. The true subgroup treatment effects are $\tau_g \in \{0.38, 0.36, 0.34, 0.32\}$ on the log-time scale. The target estimands are the subgroup-specific survival probability differences \(S_g^1(t^*)-S_g^0(t^*)\) at \(t^*=2\).}
\label{tab:results}
\resizebox{\textwidth}{!}{%
\begin{tabular}{ll rrrrrr rrrrrr}
\toprule
& & \multicolumn{6}{c}{\textbf{Gaussian Mixture}}
  & \multicolumn{6}{c}{\textbf{Gamma Mixture}} \\
\cmidrule(lr){3-8} \cmidrule(lr){9-14}
\textbf{Scenario} & \textbf{Method}
  & Rel.\ MSE & Bias & RMSE & Width & Cov. & $n_v$
  & Rel.\ MSE & Bias & RMSE & Width & Cov. & $n_v$ \\
\midrule

\multirow{10}{*}{\shortstack[l]{Decreasing\\$(0.35,0.30,$\\$0.20,0.15)$}}
& \multicolumn{13}{l}{\textit{Stratum 1 (largest, $\tau_1=0.38$)}} \\
& EMP         & 1.003 & $-0.011$ & 0.028 & 0.115 & 0.958 & 105
              & 0.885 & $-0.003$ & 0.012 & 0.053 & 0.963 & 104 \\
& BB          & 1.003 & $-0.011$ & 0.028 & 0.116 & 0.956 & 105
              & 0.885 & $-0.003$ & 0.012 & 0.053 & 0.963 & 104 \\
& HBB         & 1.000 & $-0.012$ & 0.028 & 0.113 & 0.951 & 105
              & 1.000 & 0.001    & 0.013 & 0.057 & 0.969 & 104 \\
& AFT-BART-NP & 6.110 & 0.018    & 0.069 & 0.235 & 0.906 & 105
              & 15.272 & 0.020   & 0.051 & 0.153 & 0.870 & 104 \\
\cmidrule(lr){2-14}
& \multicolumn{13}{l}{\textit{Stratum 4 (smallest, $\tau_4=0.32$)}} \\
& EMP         & 0.876 & 0.013 & 0.028 & 0.125 & 0.972 & 45
              & 2.162 & 0.013 & 0.028 & 0.115 & 0.966 & 45 \\
& BB          & 0.876 & 0.013 & 0.028 & 0.126 & 0.974 & 45
              & 2.163 & 0.013 & 0.028 & 0.116 & 0.966 & 45 \\
& HBB         & 1.000 & 0.017 & 0.030 & 0.123 & 0.952 & 45
              & 1.000 & $-0.003$ & 0.019 & 0.086 & 0.970 & 45 \\
& AFT-BART-NP & 8.490 & 0.019 & 0.087 & 0.327 & 0.933 & 45
              & 19.614 & 0.024 & 0.085 & 0.304 & 0.924 & 45 \\

\midrule

\multirow{10}{*}{\shortstack[l]{Equal\\$(0.25,0.25,$\\$0.25,0.25)$}}
& \multicolumn{13}{l}{\textit{Stratum 1 ($\tau_1=0.38$)}} \\
& EMP         & 1.006 & $-0.014$ & 0.030 & 0.124 & 0.961 & 76
              & 0.733 & $-0.003$ & 0.013 & 0.056 & 0.961 & 75 \\
& BB          & 1.006 & $-0.014$ & 0.030 & 0.125 & 0.962 & 76
              & 0.733 & $-0.003$ & 0.013 & 0.057 & 0.963 & 75 \\
& HBB         & 1.000 & $-0.015$ & 0.030 & 0.120 & 0.952 & 76
              & 1.000 & 0.004    & 0.015 & 0.064 & 0.969 & 75 \\
& AFT-BART-NP & 6.485 & 0.017    & 0.075 & 0.272 & 0.930 & 76
              & 13.755 & 0.023   & 0.055 & 0.177 & 0.904 & 75 \\
\cmidrule(lr){2-14}
& \multicolumn{13}{l}{\textit{Stratum 4 ($\tau_4=0.32$)}} \\
& EMP         & 0.893 & 0.013 & 0.027 & 0.113 & 0.971 & 75
              & 1.517 & 0.010 & 0.025 & 0.103 & 0.969 & 75 \\
& BB          & 0.893 & 0.013 & 0.027 & 0.114 & 0.972 & 75
              & 1.517 & 0.010 & 0.025 & 0.104 & 0.971 & 75 \\
& HBB         & 1.000 & 0.015 & 0.028 & 0.113 & 0.954 & 75
              & 1.000 & 0.003 & 0.020 & 0.089 & 0.972 & 75 \\
& AFT-BART-NP & 6.993 & 0.021 & 0.075 & 0.257 & 0.910 & 75
              & 13.267 & 0.018 & 0.073 & 0.240 & 0.906 & 75 \\

\midrule

\multirow{10}{*}{\shortstack[l]{Front-heavy\\$(0.38,0.38,$\\$0.12,0.12)$}}
& \multicolumn{13}{l}{\textit{Stratum 1 (large, $\tau_1=0.38$)}} \\
& EMP         & 1.006 & $-0.011$ & 0.028 & 0.114 & 0.953 & 114
              & 0.966 & $-0.003$ & 0.012 & 0.051 & 0.958 & 114 \\
& BB          & 1.006 & $-0.011$ & 0.028 & 0.114 & 0.956 & 114
              & 0.966 & $-0.003$ & 0.012 & 0.052 & 0.956 & 114 \\
& HBB         & 1.000 & $-0.011$ & 0.028 & 0.111 & 0.950 & 114
              & 1.000 & $-0.001$ & 0.012 & 0.054 & 0.967 & 114 \\
& AFT-BART-NP & 5.908 & 0.022    & 0.068 & 0.227 & 0.899 & 114
              & 15.119 & 0.019   & 0.048 & 0.148 & 0.886 & 114 \\
\cmidrule(lr){2-14}
& \multicolumn{13}{l}{\textit{Stratum 4 (small, $\tau_4=0.32$)}} \\
& EMP         & 0.928 & 0.015 & 0.030 & 0.131 & 0.972 & 36
              & 2.112 & 0.012 & 0.027 & 0.121 & 0.981 & 36 \\
& BB          & 0.929 & 0.015 & 0.030 & 0.133 & 0.973 & 36
              & 2.113 & 0.012 & 0.027 & 0.122 & 0.981 & 36 \\
& HBB         & 1.000 & 0.018 & 0.031 & 0.126 & 0.952 & 36
              & 1.000 & $-0.008$ & 0.018 & 0.081 & 0.959 & 36 \\
& AFT-BART-NP & 9.749 & 0.025 & 0.096 & 0.364 & 0.941 & 36
              & 24.961 & 0.019 & 0.092 & 0.334 & 0.931 & 36 \\

\bottomrule
\end{tabular}%
}

\end{table}

Table~1 reports results for strata 1 and 4 to contrast subgroups with different amounts and forms of covariate information. Under the decreasing and front-heavy allocations, stratum 1 is one of the largest subgroups, whereas stratum 4 is the smallest. The equal allocation provides a useful comparison in which the two strata have similar sample sizes. Their covariate distributions are the same in the Gaussian setting but differ in the Gamma setting, where stratum 4 has the more strongly skewed distribution.

EMP and BB produced nearly indistinguishable results throughout the simulation. The main comparison is therefore between these subgroup-restricted approaches and HBB, which extends the support of the covariate distribution through cross-subgroup borrowing.

In the Gaussian setting, HBB performed similarly to EMP and BB in the larger subgroups, where the subgroup-specific covariate distribution was already well represented. In the smaller subgroups, HBB generally produced slightly narrower intervals, but this gain in precision was offset by somewhat greater bias. As a result, its MSE was not lower than that of the subgroup-restricted estimators. This pattern is consistent
with the limited value of borrowing when the covariate distributions have the same, relatively simple form across subgroups.

The advantage of HBB was more apparent in the Gamma setting. For the smaller subgroups, HBB generally had lower bias and RMSE than EMP and BB, together with narrower intervals and coverage close to the nominal level. The improvement became more pronounced as the subgroup size decreased. Under the front-heavy allocation, for example, the MSE of EMP and BB in the smallest subgroup was more than twice that of HBB. In this setting, the empirical subgroup distribution contained relatively few observations from the more skewed regions of the covariate distribution. By assigning positive weight to covariate values observed elsewhere in the sample, HBB provided a more stable distribution for standardization.

The same borrowing was less useful in the larger Gamma subgroups. Here, EMP and BB generally had lower MSE than HBB, indicating that borrowing can incur a modest cost when the target subgroup already contains sufficient information about its covariate distribution. Together, the Gaussian and Gamma results suggest that the benefit of HBB is concentrated in subgroups for which the empirical covariate support is both sparse and difficult to estimate. The method offers less improvement when the subgroup distribution is already adequately represented. AFT-BART-NP had higher RMSE and substantially wider credible intervals than the three parametric estimators throughout the settings shown in Table~1. Its coverage was also generally farther from the nominal level. Although the nonparametric model allows greater flexibility in the conditional time event distribution, that flexibility was accompanied by considerably greater posterior uncertainty at the sample sizes considered here.

\subsection{Sensitivity Analysis under Informative Censoring}
\label{sec:supp_informative_censoring}

We supplemented the primary simulation with an informative censoring sensitivity analysis to assess whether the advantage of distributional borrowing persisted when incomplete follow-up depended on unobserved event risk. Unlike the primary design, an unobserved variable \(U_i\) affected both the event and censoring times, inducing dependence that remained after conditioning on the variables available to the fitted models. The true subgroup contrasts accounted for \(U_i\), whereas the same estimators used in the primary analysis were fitted without it. This design fills an evaluation gap left by the primary simulation by examining whether the benefit of regularizing the subgroup covariate distribution depends on the assumed censoring mechanism. It provides a stress test under a plausible departure from Assumption~4. Further specifications are provided in Supplementary Section~2. Table~\ref{tab:informative_main} reports performance under the same allocation and covariate distribution settings used in the primary
analysis.

\begin{table}[!htbp]
\centering
\caption{Simulation results under informative censoring across three subgroup allocation scenarios and two covariate settings. Results are based on $S=1{,}000$ replications with $N=300$ subjects and $K=4$ subgroups. Strata~1 and~4 represent the largest and smallest subgroups under the decreasing and front-heavy scenarios, respectively. The true subgroup treatment effects are $\tau_g \in \{0.38, 0.36, 0.34, 0.32\}$ on the log-time scale. The target estimands are the subgroup-specific survival probability differences \(S_g^1(t^*)-S_g^0(t^*)\) at \(t^*=2\).}
\label{tab:informative_main}
\resizebox{\textwidth}{!}{%
\begin{tabular}{ll rrrrrr rrrrrr}
\toprule
& & \multicolumn{6}{c}{\textbf{Gaussian Mixture}}
  & \multicolumn{6}{c}{\textbf{Gamma Mixture}} \\
\cmidrule(lr){3-8} \cmidrule(lr){9-14}
\textbf{Scenario} & \textbf{Method}
  & Rel.\ MSE & Bias & RMSE & Width & Cov. & $n_v$
  & Rel.\ MSE & Bias & RMSE & Width & Cov. & $n_v$ \\
\midrule

\multirow{10}{*}{\shortstack[l]{Decreasing\\$(0.35,0.30,$\\$0.20,0.15)$}}
& \multicolumn{13}{l}{\textit{Stratum 1 (largest, $\tau_1=0.38$)}} \\
& EMP         & 1.013 & $-0.007$ & 0.028 & 0.122 & 0.968 & 105
              & 0.813 & 0.0002    & 0.013 & 0.057 & 0.975 & 104 \\
& BB          & 1.013 & $-0.007$ & 0.028 & 0.123 & 0.966 & 105
              & 0.813 & 0.0002    & 0.013 & 0.058 & 0.975 & 104 \\
& HBB         & 1.000 & $-0.007$ & 0.028 & 0.119 & 0.966 & 105
              & 1.000 & 0.004    & 0.015 & 0.060 & 0.968 & 104 \\
& AFT-BART-NP & 6.118 & 0.014    & 0.068 & 0.243 & 0.923 & 105
              & 11.768 & 0.014   & 0.050 & 0.157 & 0.880 & 104 \\
\cmidrule(lr){2-14}
& \multicolumn{13}{l}{\textit{Stratum 4 (smallest, $\tau_4=0.32$)}} \\
& EMP         & 0.871 & 0.017 & 0.031 & 0.141 & 0.975 & 45
              & 2.292 & 0.015 & 0.030 & 0.125 & 0.964 & 45 \\
& BB          & 0.871 & 0.017 & 0.031 & 0.143 & 0.976 & 45
              & 2.292 & 0.016 & 0.030 & 0.126 & 0.964 & 45 \\
& HBB         & 1.000 & 0.021 & 0.034 & 0.140 & 0.961 & 45
              & 1.000 & $-0.000037$ & 0.020 & 0.095 & 0.981 & 45 \\
& AFT-BART-NP & 7.964 & 0.028 & 0.095 & 0.355 & 0.934 & 45
              & 22.226 & 0.031 & 0.093 & 0.327 & 0.919 & 45 \\

\midrule

\multirow{10}{*}{\shortstack[l]{Equal\\$(0.25,0.25,$\\$0.25,0.25)$}}
& \multicolumn{13}{l}{\textit{Stratum 1 ($\tau_1=0.38$)}} \\
& EMP         & 1.033 & $-0.009$ & 0.030 & 0.133 & 0.973 & 76
              & 0.651 & $−0.0000479$    & 0.014 & 0.062 & 0.979 & 75 \\
& BB          & 1.033 & $-0.009$ & 0.030 & 0.134 & 0.973 & 76
              & 0.651 & $−0.0000454$    & 0.014 & 0.062 & 0.981 & 75 \\
& HBB         & 1.000 & $-0.010$ & 0.030 & 0.127 & 0.968 & 76
              & 1.000 & 0.007    & 0.017 & 0.069 & 0.957 & 75 \\
& AFT-BART-NP & 6.496 & 0.013    & 0.075 & 0.282 & 0.936 & 76
              & 10.008 & 0.016   & 0.053 & 0.183 & 0.912 & 75 \\
\cmidrule(lr){2-14}
& \multicolumn{13}{l}{\textit{Stratum 4 ($\tau_4=0.32$)}} \\
& EMP         & 0.877 & 0.016 & 0.030 & 0.129 & 0.957 & 75
              & 1.528 & 0.013 & 0.026 & 0.113 & 0.974 & 75 \\
& BB          & 0.877 & 0.016 & 0.030 & 0.130 & 0.960 & 75
              & 1.528 & 0.013 & 0.026 & 0.113 & 0.973 & 75 \\
& HBB         & 1.000 & 0.019 & 0.032 & 0.129 & 0.941 & 75
              & 1.000 & 0.006 & 0.021 & 0.099 & 0.984 & 75 \\
& AFT-BART-NP & 6.302 & 0.029 & 0.081 & 0.278 & 0.911 & 75
              & 13.453 & 0.024 & 0.079 & 0.257 & 0.889 & 75 \\

\midrule

\multirow{10}{*}{\shortstack[l]{Front-heavy\\$(0.38,0.38,$\\$0.12,0.12)$}}
& \multicolumn{13}{l}{\textit{Stratum 1 (large, $\tau_1=0.38$)}} \\
& EMP         & 1.019 & $-0.006$ & 0.028 & 0.120 & 0.953 & 114
              & 0.904 & $-0.001$ & 0.012 & 0.055 & 0.969 & 114 \\
& BB          & 1.018 & $-0.006$ & 0.028 & 0.121 & 0.956 & 114
              & 0.905 & $-0.001$ & 0.013 & 0.056 & 0.967 & 114 \\
& HBB         & 1.000 & $-0.007$ & 0.028 & 0.118 & 0.954 & 114
              & 1.000 & 0.002    & 0.013 & 0.057 & 0.970 & 114 \\
& AFT-BART-NP & 5.898 & 0.019    & 0.068 & 0.233 & 0.918 & 114
              & 12.624 & 0.013   & 0.047 & 0.152 & 0.901 & 114 \\
\cmidrule(lr){2-14}
& \multicolumn{13}{l}{\textit{Stratum 4 (small, $\tau_4=0.32$)}} \\
& EMP         & 0.913 & 0.017 & 0.032 & 0.148 & 0.982 & 36
              & 2.373 & 0.014 & 0.028 & 0.131 & 0.983 & 36 \\
& BB          & 0.913 & 0.017 & 0.032 & 0.150 & 0.981 & 36
              & 2.375 & 0.014 & 0.028 & 0.132 & 0.983 & 36 \\
& HBB         & 1.000 & 0.021 & 0.034 & 0.145 & 0.964 & 36
              & 1.000 & $-0.006$ & 0.018 & 0.091 & 0.971 & 36 \\
& AFT-BART-NP & 9.373 & 0.033 & 0.103 & 0.394 & 0.938 & 36
              & 29.504 & 0.025 & 0.100 & 0.359 & 0.916 & 36 \\

\bottomrule
\end{tabular}%
}

\end{table}


The relative performance of the estimators depended more strongly on the covariate distribution than on subgroup size alone. Under the Gaussian design, HBB performed similarly to the subgroup-restricted
estimators and did not reduce MSE, including in stratum 4. The same was true for the larger Gamma subgroup, where the empirical covariate distribution was already adequately represented.

A different pattern emerged for stratum 4 under the Gamma design. Here, HBB consistently reduced absolute bias and RMSE and produced narrower intervals across all three allocation schemes. The MSE of EMP and BB ranged from approximately 1.5 to 2.4 times that of HBB, with the largest difference under the front-heavy allocation. The advantage remained under equal allocation, indicating that it reflected the difficulty of representing the skewed covariate distribution rather than small subgroup size alone. EMP and BB otherwise remained nearly indistinguishable, while AFT-BART-NP produced higher RMSE and wider intervals throughout the settings considered. Coverage for the parametric estimators generally remained close to the nominal level. Compared with the primary analysis, informative censoring generally increased RMSE and interval width but did not completely change where HBB provided an advantage. The sensitivity analysis shows that the improvement from distributional regularization in the more challenging subgroup was not confined to the conditional independent censoring design, while the
deterioration in absolute performance clarifies the remaining need for methods that explicitly address informative censoring.

\section{Application to the PSC study}\label{sec4}
\label{sec:psc_application}

The data used in this study were derived from the Pediatric PSC Consortium, an international research registry involving multiple sites across North America, Europe, and other regions\cite{deneau2017natural}. The registry includes patients diagnosed with primary sclerosing cholangitis (PSC) with disease onset before 18 years of age. Clinical data were collected through detailed medical record review at each participating site, including laboratory measurements, imaging results, and treatment information.

PSC diagnosis was established based on cholestatic biochemical profiles together with characteristic imaging or histopathological findings, consistent with established clinical criteria. Coexisting conditions, including inflammatory bowel disease (IBD) and features of autoimmune hepatitis (AIH) were recorded through medical chart review. Pharmacy data were also collected, including treatment with ursodiol, concurrent vancomycin use, and other relevant medications. For the present analysis, we included 260 patients with complete data on baseline covariates, treatment status, and longitudinal follow-up; 182 received ursodiol and 78 did not.

In PSC, gamma-glutamyl transferase (GGT) is a widely used biomarker of bile duct injury and disease activity, and reductions in GGT have been associated with improved clinical outcomes in prior studies \cite{deneau2021oral}. The Pediatric PSC Consortium data have previously been used to compare ursodiol, vancomycin, and no treatment through propensity-score matching followed by analyses within clinically defined risk strata \cite{deneau2021oral}. The present study addresses the same underlying
clinical question from a different analytic perspective. Rather than forming matched treatment groups and subsequently comparing outcomes within separate strata, we estimate treatment-specific survival probabilities under ursodiol and no ursodiol across the prespecified age and SCOPE risk subgroups. The analysis incorporates right-censored
follow-up, adjusts for concurrent vancomycin use and other baseline covariates, and standardizes the conditional survival estimates to the covariate distribution of each subgroup. It complements the earlier matched analysis by characterizing how the estimated treatment contrast evolves over time within the clinically defined subgroups.

Motivated by this clinical relevance and prior work within this cohort, we defined the primary outcome as time to GGT normalization \cite{deneau2018gamma}($<$ 50 U/L). In this study, subgroup specific effects were evaluated across prespecified age groups and baseline SCOPE risk strata. Age was categorized as 0--10 years, 11--15 years, and older than 15 years. Baseline disease severity was summarized using the Sclerosing Cholangitis Outcomes in Pediatrics (SCOPE) index \cite{deneau2021oral}, a validated prognostic tool that assigns an integer score from 0 to 11 based on five clinical parameters: total bilirubin, serum albumin, platelet count, GGT, and the presence or absence of large-duct cholangiographic involvement. SCOPE scores of 0--3, 4--5, and 6--11 were categorized as low-, medium-, and high-risk groups respectively. The adjusted analysis included baseline IBD status, PSC-AIH overlap, baseline log-transformed GGT, and concurrent vancomycin use, since vancomycin is used in a subset of pediatric PSC patients and may independently influence biochemical trajectories.

Treatment with ursodiol was associated with an increase in the probability of GGT normalization across the prespecified age and SCOPE risk subgroups, with a clear time-dependent pattern. At 1 month, the estimated risk differences were smaller, while larger effects were observed by 6 months and remained relatively stable at 9 months. This trend was consistent across both the overall estimates and subgroup specific results.

Across age groups and SCOPE risk strata, although the magnitude varies modestly between subgroups, the direction of the effect remains positive. As shown in Figure Supplementary~S1, most subgroup estimates increase from 1 to 6 months, with limited additional change between 6 and 9 months. The uncertainty intervals overlap substantially across subgroups and all extend below zero, indicating that while the direction of the effect is consistently positive, estimates carry considerable uncertainty, particularly in smaller subgroups. The wide credible intervals likely reflect limited subgroup sample sizes and uncertainty associated with adjustment for baseline covariates.

When SCOPE risk was incorporated as a heterogeneity variable, high-risk patients tended to have slightly larger estimated benefits within each age group, whereas low-risk patients generally showed smaller effects. However, these differences were modest, and the credible intervals overlapped substantially across subgroups. Therefore, the current data do not provide strong evidence that treatment response differs meaningfully by age group or SCOPE risk category. Overall, the estimated benefit of ursodiol became more apparent by 6 months and remained relatively stable thereafter, with a consistently positive direction across subgroups.

\begin{table}[!htbp]
\centering
\caption{Baseline characteristics by ursodiol treatment group}
\label{tab:baseline}
\small
\begin{tabular}{lcccc}
\hline
\textbf{Variable} 
& \textbf{Overall (n = 260)} 
& \textbf{No Urso (n = 78)} 
& \textbf{Urso (n = 182)} 
& \textbf{$p$-value} \\
\hline

\textbf{Sex} & & & & 0.444 \\
\quad Female & 111 (43\%) & 30 (38\%) & 81 (45\%) & \\
\quad Male   & 149 (57\%) & 48 (62\%) & 101 (55\%) & \\

\textbf{IBD} & 192 (74\%) & 69 (88\%) & 123 (68\%) & $<0.001$ \\

\textbf{Age group} & & & & $<0.001$ \\
\quad 0--10  & 63 (24\%)  & 7 (9.0\%) & 56 (31\%) & \\
\quad 11--15     & 100 (38\%) & 36 (46\%) & 64 (35\%) & \\
\quad 15+   & 97 (37\%)  & 35 (45\%) & 62 (34\%) & \\

\textbf{SCOPE risk} & & & & 0.019 \\
\quad Low    & 99 (38\%)  & 30 (38\%) & 69 (38\%) & \\
\quad Medium & 101 (39\%) & 38 (49\%) & 63 (35\%) & \\
\quad High   & 60 (23\%)  & 10 (13\%) & 50 (27\%) & \\

\textbf{SCOPE index} 
& 4 [3, 5] 
& 4 [3, 5] 
& 4 [3, 6] 
& 0.091 \\

\textbf{Vancomycin (concurrent)} 
& 43 (17\%) 
& 21 (27\%) 
& 22 (12\%) 
& 0.006 \\

\textbf{GGT (U/L)} 
& 204 [97, 393] 
& 205 [101, 425] 
& 204 [96, 375] 
& 0.870 \\

\textbf{Bilirubin (mg/dL)} 
& 0.60 [0.40, 1.21] 
& 0.50 [0.30, 0.90] 
& 0.65 [0.40, 1.50] 
& 0.007 \\

\textbf{ALT (U/L)} 
& 100.50 [48, 213.50] 
& 100.50 [41, 221] 
& 101.50 [49, 208] 
& 0.905 \\

\textbf{ALP (U/L)} 
& 305.50 [196, 516] 
& 293 [159, 470] 
& 316 [207, 535] 
& 0.273 \\

\textbf{Platelet ($\times 10^3/\mu$L)} 
& 302.50 [225, 405] 
& 327.50 [260, 430] 
& 292 [221, 388] 
& 0.062 \\

\textbf{Follow-up time (years)} 
& 2.29 [0.49, 4.70] 
& 2.14 [0.67, 3.59] 
& 2.37 [0.30, 5.60] 
& 0.292 \\

\textbf{GGT normalization} 
& 74 (28\%) 
& 9 (12\%) 
& 65 (36\%) 
& $<0.001$ \\

\hline
\multicolumn{5}{l}{\footnotesize Values are presented as n (\%) or median [Q1, Q3].} \\
\multicolumn{5}{l}{\footnotesize $p$-values are from Pearson's chi-squared test or Wilcoxon rank-sum test.} \\
\end{tabular}
\end{table}

\begin{figure}[!htbp]
\centering
\includegraphics[width=0.75\textwidth]{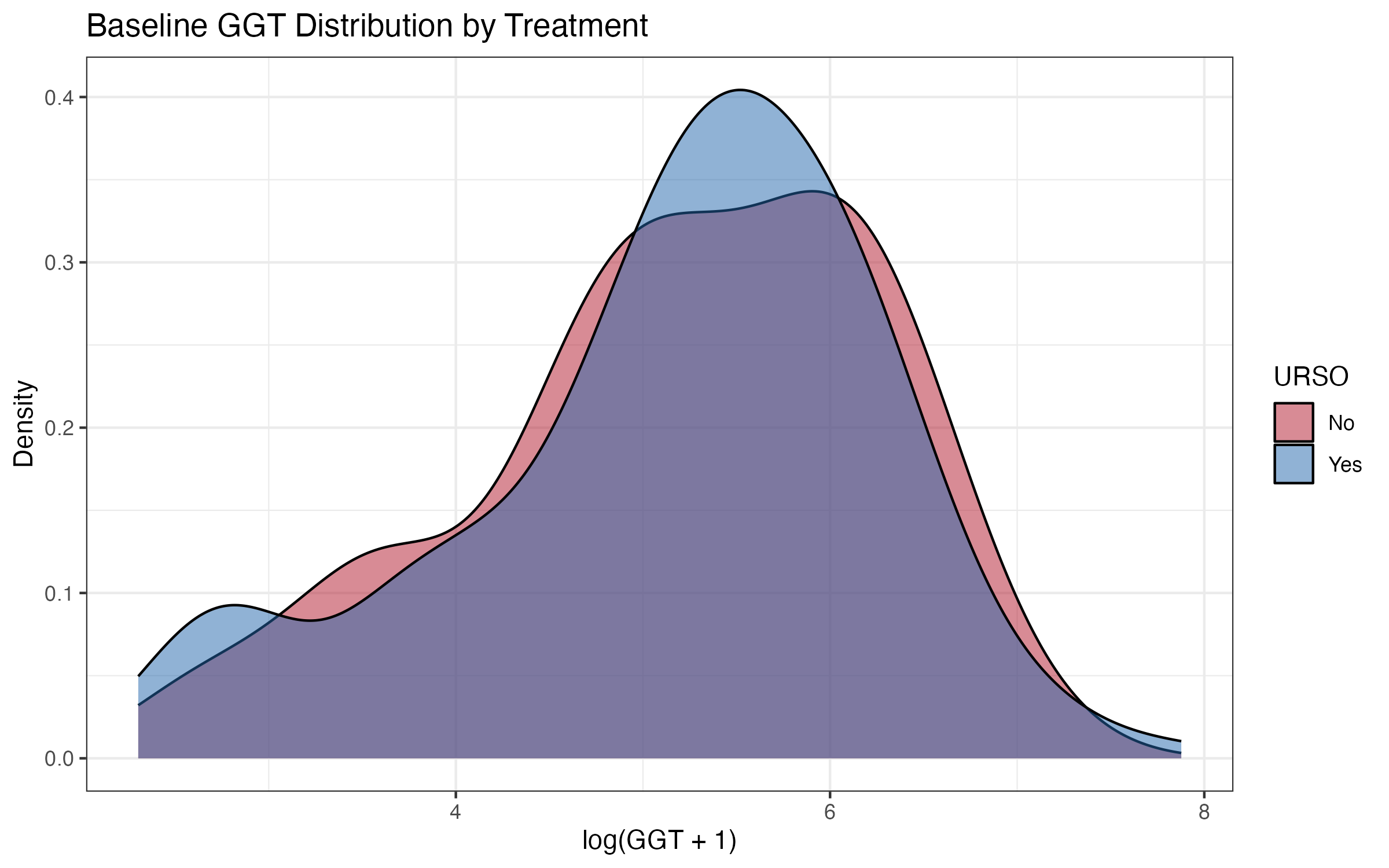}
\caption{Baseline GGT distribution by ursodiol treatment group. Density curves are shown on the
log$(GGT + 1)$ scale for the treated (URSO, blue) and untreated (No URSO, pink) groups.}
\label{fig:ggt_distribution}
\end{figure}

Baseline GGT distributions differed between the ursodiol-treated and untreated groups (Figure~\ref{fig:ggt_distribution}). The treated group showed a slightly higher peak density at moderate GGT values and a secondary mass at lower values, while the untreated group exhibited a broader right tail. This distributional difference motivates the covariate-adjusted framework used in the primary analysis, as direct comparisons of treatment outcomes without adjustment for baseline GGT and other prognostic factors could be misleading.

The unadjusted Kaplan--Meier curves showed earlier observed GGT normalization among patients who received ursodiol than among untreated patients (Supplementary Figure~S4). The adjusted posterior survival curves in Figure~3 showed a similar direction across the prespecified age and SCOPE risk groups. Because these curves represent the probability of remaining without GGT normalization, the lower curves under ursodiol correspond to a greater cumulative probability of normalization during follow-up.

\begin{figure}[!htbp]
\centering
\includegraphics[width=0.75\textwidth]{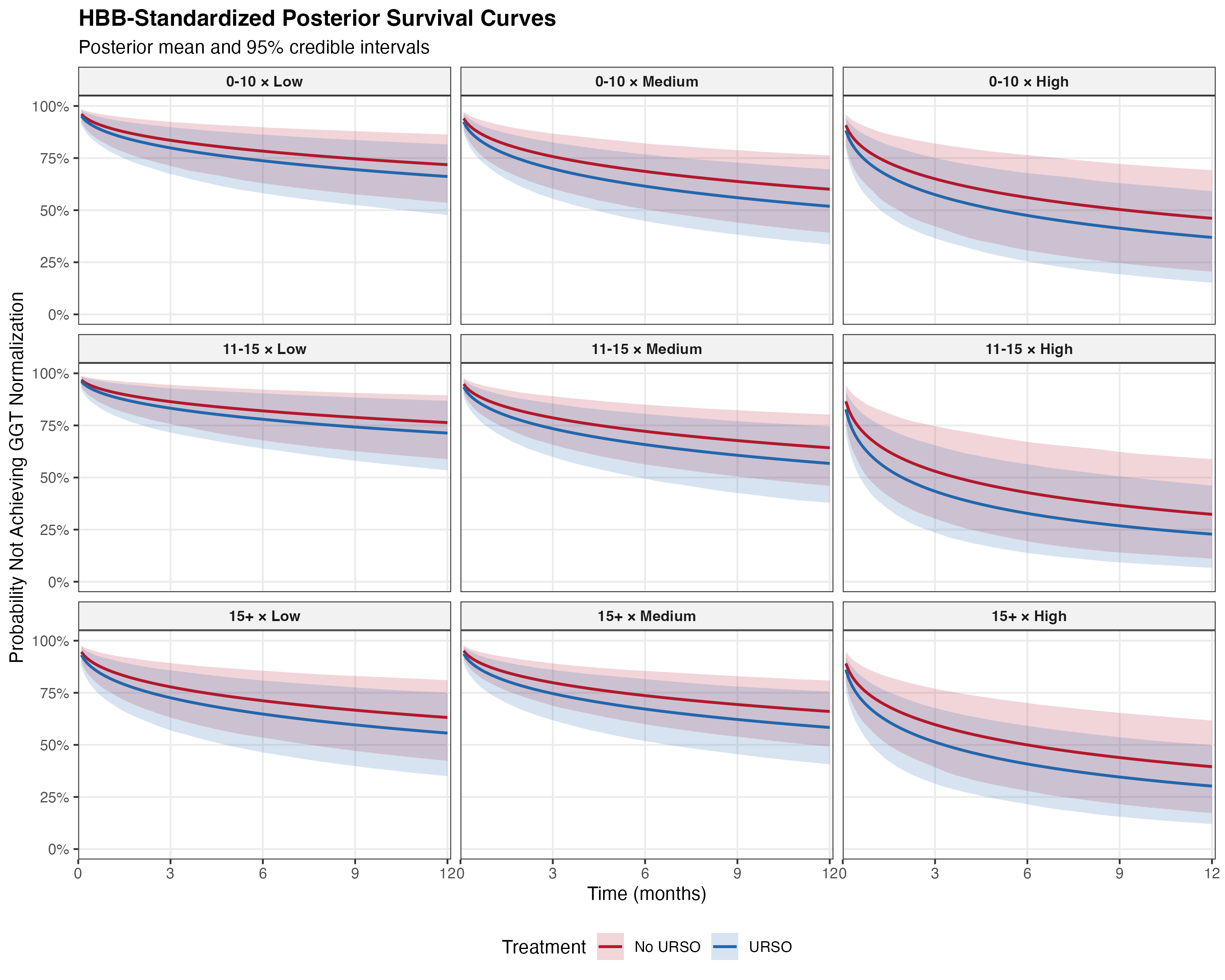}
\caption{HBB-standardized posterior survival curves under URSO and no URSO}
\label{fig:standardized posterior survival curves}
\end{figure}

On the cumulative probability scale, the estimated absolute benefit of ursodiol was relatively small at 1 month, generally increased by 6 months, and changed little between 6 and 9 months (Table~3). At 6 months, ursodiol was associated with an approximately 4 to 10 percentage point increase in the probability of GGT normalization across the subgroups. Clinically, this pattern suggests that the difference in biochemical response emerged mainly during the first
several months of treatment and was subsequently maintained, rather than continuing to widen later in follow-up. The continuous trajectories in Supplementary Figure~S1 showed the same early increase and subsequent flattening.

The posterior mean contrasts favored ursodiol in every age and SCOPE risk group, with posterior probabilities of benefit ranging from approximately 0.88 to 0.93. Within each age group, the estimates tended to be larger among patients with higher baseline SCOPE risk, suggesting that patients with more severe disease may experience a somewhat greater absolute improvement in GGT normalization. However, the
differences between risk groups were modest relative to the uncertainty in their estimated magnitude. The results therefore support a broadly consistent direction of benefit across the prespecified subgroups more strongly than they support a distinct response for any single age or SCOPE risk category. Posterior probabilities over the full follow-up period are presented in Supplementary Figure~S2. Using this framework, treatment effects are summarized through absolute differences in the probability of GGT normalization. 

\begin{table}[htbp]
\centering
\caption{Subgroup-specific posterior g-computation estimates of
absolute treatment benefit under ursodiol versus no ursodiol.}
\label{tab:subgroup_rd}
\small
\begin{tabular}{llcccc}
\hline
Age group
& SCOPE risk
& Time (months)
& Risk difference
& 95\% CrI
& \(\Pr(\mathrm{RD}>0)\) [95\% interval] \\
\hline

0--10 & Low
& 1 & 0.024 & [$-$0.020, 0.068] & 0.892 [0.873, 0.911] \\
& & \textbf{6}
& \textbf{0.046}
& \textbf{[$-$0.036, 0.130]}
& \textbf{0.892 [0.872, 0.912]} \\
& & 9 & 0.052 & [$-$0.040, 0.145] & 0.892 [0.872, 0.910] \\
\hline

0--10 & Medium
& 1 & 0.041 & [$-$0.019, 0.110] & 0.907 [0.890, 0.925] \\
& & \textbf{6}
& \textbf{0.071}
& \textbf{[$-$0.032, 0.182]}
& \textbf{0.907 [0.889, 0.924]} \\
& & 9 & 0.078 & [$-$0.035, 0.197] & 0.907 [0.889, 0.925] \\
\hline

0--10 & High
& 1 & 0.057 & [$-$0.031, 0.148] & 0.902 [0.884, 0.920] \\
& & \textbf{6}
& \textbf{0.086}
& \textbf{[$-$0.044, 0.224]}
& \textbf{0.902 [0.882, 0.920]} \\
& & 9 & 0.090 & [$-$0.047, 0.237] & 0.902 [0.882, 0.919] \\
\hline

11--15 & Low
& 1 & 0.021 & [$-$0.018, 0.062] & 0.883 [0.864, 0.904] \\
& & \textbf{6}
& \textbf{0.040}
& \textbf{[$-$0.033, 0.120]}
& \textbf{0.883 [0.862, 0.903]} \\
& & 9 & 0.046 & [$-$0.039, 0.136] & 0.883 [0.864, 0.901] \\
\hline

11--15 & Medium
& 1 & 0.035 & [$-$0.021, 0.093] & 0.914 [0.896, 0.930] \\
& & \textbf{6}
& \textbf{0.064}
& \textbf{[$-$0.034, 0.160]}
& \textbf{0.914 [0.897, 0.930]} \\
& & 9 & 0.071 & [$-$0.037, 0.177] & 0.914 [0.897, 0.930] \\
\hline

11--15 & High
& 1 & 0.080 & [$-$0.032, 0.197] & 0.926 [0.909, 0.942] \\
& & \textbf{6}
& \textbf{0.100}
& \textbf{[$-$0.036, 0.251]}
& \textbf{0.926 [0.910, 0.942]} \\
& & 9 & 0.098 & [$-$0.036, 0.261] & 0.926 [0.909, 0.941] \\
\hline

15+ & Low
& 1 & 0.036 & [$-$0.019, 0.098] & 0.896 [0.876, 0.915] \\
& & \textbf{6}
& \textbf{0.064}
& \textbf{[$-$0.033, 0.171]}
& \textbf{0.896 [0.878, 0.914]} \\
& & 9 & 0.070 & [$-$0.037, 0.190] & 0.896 [0.877, 0.915] \\
\hline

15+ & Medium
& 1 & 0.036 & [$-$0.014, 0.098] & 0.913 [0.894, 0.930] \\
& & \textbf{6}
& \textbf{0.065}
& \textbf{[$-$0.026, 0.168]}
& \textbf{0.913 [0.895, 0.929]} \\
& & 9 & 0.072 & [$-$0.030, 0.187] & 0.913 [0.896, 0.930] \\
\hline

15+ & High
& 1 & 0.065 & [$-$0.029, 0.164] & 0.923 [0.906, 0.940] \\
& & \textbf{6}
& \textbf{0.092}
& \textbf{[$-$0.039, 0.240]}
& \textbf{0.923 [0.906, 0.940]} \\
& & 9 & 0.093 & [$-$0.039, 0.241] & 0.923 [0.906, 0.940] \\
\hline
\end{tabular}
\end{table}

These findings extend earlier stratified analyses by describing the ursodiol contrast across age and SCOPE risk groups within an adjusted causal survival framework. Although the posterior estimates showed a generally positive direction and a modest gradient toward greater benefit in higher-risk groups, uncertainty in the relative magnitude of these subgroup effects remains substantial. The current evidence offers limited support for distinguishing treatment response by age or SCOPE risk and does not justify subgroup-specific recommendations for ursodiol use. Further evaluation in larger, prospectively collected cohorts will be important for assessing the reproducibility of the observed severity gradient and separating the contributions of ursodiol from those of co-administered therapies.

\section{Discussion}
\label{sec:discussion}

We developed an extension of the hierarchical Bayesian bootstrap for population-level subgroup causal inference with right censored time-to-event outcomes. The proposed framework combines a conditional AFT survival model with hierarchical regularization of the subgroup covariate distributions used for g-formula standardization that propagating uncertainty from the conditional event-time distribution and the subgroup-specific standardization distribution into subgroup-specific survival probabilities and causal contrasts. It is complementary to existing flexible Bayesian survival models, which primarily regularize the conditional event-time response surface \citep{Arman2024,Henderson2020-vb}.

The simulation results suggest that HBB is most useful when a target subgroup's covariate distribution is sparsely or irregularly represented and the other subgroups provide commensurate covariate information. Little improvement should be expected when the empirical subgroup distribution is already well represented, and incompatible borrowing may introduce bias. The clearest gains of HBB occurred in the smaller subgroup under the Gamma simulated covariates distribution, where the skewed covariate distribution was difficult to represent using the target subgroup observations alone. HBB reduced estimation error by assigning positive posterior mass to covariate profiles observed in other subgroups. The persistence of this improvement under equal subgroup allocation suggests that the value of borrowing was determined not only by subgroup size but also by the adequacy of empirical covariate support. Nevertheless, reductions in interval width should be interpreted as improved precision only when accompanied by adequate empirical coverage. EMP and BB yielded similar results because both standardized over covariate profiles observed within the target subgroup; the additional random weighting under BB contributed relatively little uncertainty compared with the event-time model in the settings considered. HBB also performed similarly to EMP and BB when the subgroup covariate distribution was adequately represented, as in stratum 1 and under the Gaussian design. In these settings, additional borrowing provided little benefit and could result in a modest efficiency cost.

AFT-BART-NP produced greater estimation error and wider credible intervals in the scenarios considered. This result does not establish a general advantage of parametric AFT models. Flexible Bayesian AFT models remain valuable when nonlinear covariate effects, treatment--covariate interactions, or departures from a parametric error distribution are plausible \citep{Henderson2020-vb}. Their wider intervals may reflect the uncertainty associated with this additional flexibility, particularly at the available sample sizes. The comparison may also favor the parametric model when the simulated event time mechanism is close to its assumptions.

In our setting, borrowing across subgroups requires partial exchangeability of their covariate distributions. The current borrowing structure does not distinguish among potential source subgroups according to their similarity to the target subgroup. In the current construction, the concentration parameter \(\alpha_g\) controls the overall amount of information borrowed by subgroup \(g\), while the shared distribution \(F_0\) determines where that information comes from. Because \(F_0\) is constructed from the pooled sample, the contribution of a source subgroup is driven primarily by its sample size rather than its covariate similarity to the target. In the proposed work, a fixed calibration of \(\alpha_g=nM/n_g\) is used for the concentration parameter. A full Bayesian specification is possible by assigning a hyperprior to the reference size \(M\), or directly to the pooling fraction \(\lambda_g=\alpha_g/(\alpha_g+n_g)\), and propagating uncertainty in the pooling strength through the posterior g-formula. Gamma hyperpriors and auxiliary variable updates for concentration parameters are well established for Dirichlet-process and hierarchical Dirichlet-process models \citep{EscobarWest1995,Teh2006}. Furthermore, borrowing across subgroups can reduce variance at the cost of biasing the finite-sample estimate toward an incompatible distribution. This consideration motivates routine borrowing diagnostics to assess borrowing compatibility. Applied analyses should consider assessing the posterior amount of information borrowed by each target subgroup, the contribution of each source subgroup, sensitivity to the concentration parameters, and results under a no-borrowing specification. A more adaptive extension could replace the single pooled base distribution with a target-specific mixture $F_{0g}=\sum_{h\ne g}\omega_{gh}F_h$, where the weights depend on jointly estimated similarities between the target and source covariate distributions. Distributional-discrepancy and robust dynamic-borrowing methods provide possible extensions for favoring commensurate sources and discounting conflicting information \citep{zheng2022borrowing,jiang2023elastic}.

One limitation of the proposed work is the reliance on the conditional AFT model. Distributional borrowing cannot correct misspecification of \(S_a(t\mid x,g)\). Posterior predictive checks and sensitivity analyses under alternative event-time distributions should performed. Flexible Bayesian AFT models is a potential approach but as we have shown in the simulation study, it may be inefficient with small sample size \citep{Henderson2020-vb}. Another direction is to combine HBB standardization with an augmented estimator involving the treatment and censoring mechanisms. Doubly robust estimators of treatment-specific survival distributions provide a foundation for such an extension \citep{bai2013doubly,schnitzer2016double}. The present likelihood also relies on conditionally independent censoring. If failure and censoring remain dependent after conditioning on observed information, identification requires additional assumptions, such as a joint event--censoring model or a sensitivity parameter governing their residual dependence. In the presence of competing events, an extension could standardize treatment-specific cumulative incidence functions  \citep{stensrud2022separable,xu2022bayesian}.

Future work could develop similarity adaptive borrowing structures and combine HBB with more flexible time event models when the AFT specification is uncertain. Extensions to informative censoring, competing risks, and time-varying treatments would require additional models for those processes while retaining the HBB component for the covariate distributions used in causal standardization.

\clearpage
\bibliographystyle{unsrt}
\bibliography{reference}

\section*{Supplementary Materials}

\subsection*{Conditional independence factorization}\label{app:cindep}

Under the assumption that the potential event time $T_i^\ast$ and censoring time $C_i$ are conditionally independent given $(A_i, X_i, G_i)$, the observed survival function factorizes as
\begin{equation}\label{eq:cindep_factorization}
\begin{aligned}
\Pr(Y_i \ge t \mid A_i=a, X_i=x, G_i=g)
&= \Pr(T_i^\ast \ge t,\; C_i \ge t \mid A_i=a, X_i=x, G_i=g) \\
&= \Pr(T_i^\ast \ge t \mid A_i=a, X_i=x, G_i=g) \\
&\quad \times \Pr(C_i \ge t \mid A_i=a, X_i=x, G_i=g).
\end{aligned}
\end{equation}

\subsection*{HBB posterior weights}\label{app:hbb_weights}

Given the hierarchical Bayesian bootstrap prior $F_g \mid \alpha_g, F_0 \sim \mathrm{DP}(\alpha_g F_0 + \sum_{i \in S_g} \delta_{X_i})$ with $F_0(\cdot) = \sum_{i=1}^n \pi_i \, \delta_{X_i}(\cdot)$, the posterior mean of $F_g$ is
\begin{equation}\label{eq:hbb_posterior_mean}
\begin{aligned}
\mathbb{E}\!\left\{F_g(\cdot) \mid \pi, \alpha_g, X_g\right\}
&= \frac{\alpha_g}{\alpha_g + n_g}\, F_0(\cdot)
   + \frac{1}{\alpha_g + n_g} \sum_{i \in S_g} \delta_{X_i}(\cdot) \\
&= \frac{\alpha_g}{\alpha_g + n_g} \sum_{i=1}^n \pi_i\, \delta_{X_i}(\cdot)
   + \frac{1}{\alpha_g + n_g} \sum_{i \in S_g} \delta_{X_i}(\cdot) \\
&= \frac{1}{\alpha_g + n_g}
   \left\{
     \sum_{i \notin S_g} \alpha_g \pi_i\, \delta_{X_i}(\cdot)
     + \sum_{i \in S_g} (\alpha_g \pi_i + 1)\, \delta_{X_i}(\cdot)
   \right\} \\
&= \sum_{i=1}^n
   \underbrace{\frac{\alpha_g \pi_i + \mathbb{I}(i \in S_g)}{\alpha_g + n_g}}_{\displaystyle =:\; \omega_{gi}}
   \; \delta_{X_i}(\cdot).
\end{aligned}
\end{equation}
Equation~(2) gives the posterior mean weight assigned to each observed covariate value. Posterior computation uses the corresponding random Dirichlet weights rather than fixing the weights at these posterior means, thereby propagating uncertainty in \(F_g\) through the standardization step.

\subsection*{Posterior g-computation}\label{app:gcomp}
For each posterior draw, combining the sampled HBB weights from~\eqref{eq:hbb_posterior_mean}  with the conditional survival function gives

\begin{equation}\label{eq:posterior_gcomputation}
\begin{aligned}
S_g^a(t)
&= \int S_{T^\ast}(t \mid A=a, X=x, G=g)\, dF_g(x) \\
&= \sum_{i=1}^n \omega_{gi}\, S_{T^\ast}(t \mid a, X_i, g) \\
&= \sum_{i=1}^n \omega_{gi}
   \left[1 - F_{\varepsilon,\theta}\!\left(\log t - m_\theta(a, X_i, g)\right)\right].
\end{aligned}
\end{equation}

\subsection*{Proof of the g-formula identification}\label{app:proof_gformula}

By the law of total expectation,
\begin{align}
S_g^a(t)
&= \Pr\!\big(T_i^*(a)>t \mid G_i=g\big) \nonumber\\
&= \mathbb{E}\!\left[
\Pr\!\big(T_i^*(a)>t \mid X_i, G_i=g\big)
\ \middle|\ G_i=g
\right].
\label{eq:step_totalexp}
\end{align}
By conditional exchangeability (Assumption~2) and positivity
(Assumption~3),
\begin{equation}
\Pr\!\big(T_i^*(a)>t \mid X_i, G_i=g\big)
=
\Pr\!\big(T_i^*(a)>t \mid A_i=a, X_i, G_i=g\big).
\label{eq:step_exchangeability}
\end{equation}
By consistency (Assumption~1),
\begin{equation}
\Pr\!\big(T_i^*(a)>t \mid A_i=a, X_i, G_i=g\big)
=
\Pr\!\big(T_i^*>t \mid A_i=a, X_i, G_i=g\big).
\label{eq:step_consistency}
\end{equation}

Substituting~\eqref{eq:step_exchangeability}--\eqref{eq:step_consistency} into~\eqref{eq:step_totalexp} yields the identification formula used for subgroup standardization. \hfill $\square$

These derivations establish the posterior computation used in the main analysis. The next section examines how the estimators behave when the conditional independent censoring assumption is violated.

\subsection*{Additional specifications for the informative censoring sensitivity analysis}
\label{app:informative_details}

The main text reports the motivation and results of the informative censoring sensitivity analysis. This section provides additional details of the data-generating and estimation procedures. All simulation components not described below were unchanged from the
primary analysis.

The shared latent variable was generated as \(U_i\sim N(0,1)\), independently of treatment, the observed covariates, and subgroup membership. It entered the event-time and censoring-time linear predictors with coefficients 0.15 and 0.80, respectively. Treatment
shifted the censoring-time linear predictor by
\((-0.50,-0.75,-1.00,-1.50)\) across the four subgroups. Thus, the dependence remaining after adjustment arose from the shared \(U_i\), rather than from the observed treatment or subgroup effects on
censoring.

The true subgroup survival contrasts were evaluated from the data-generating event-time model by averaging over both the subgroup covariate distribution and the distribution of \(U_i\). Including \(U_i\) in this calculation ensures that the simulation truth corresponds to the marginal subgroup estimand, even though \(U_i\) is unavailable to the fitted models.

A Weibull AFT model was fitted to each of the event and censoring processes. For an observed event, the likelihood contained the event-time density and the probability of remaining uncensored to the
observed time; for a censored observation, it contained the event-time survival probability and the censoring-time density. Using the notation
introduced in the main text, the contribution was
\[
L_i
\propto
\left\{
f_{T^*}(Y_i\mid A_i,X_i,G_i)
S_C(Y_i\mid A_i,X_i,G_i)
\right\}^{\Delta_i}
\left\{
S_{T^*}(Y_i\mid A_i,X_i,G_i)
f_C(Y_i\mid A_i,X_i,G_i)
\right\}^{1-\Delta_i}.
\]
Although both likelihood components were estimated, neither included \(U_i\), and they were not connected through a shared latent structure. The fitted likelihood therefore described the marginal event and
censoring distributions without accounting for their residual dependence. The HBB reference size was fixed at \(M=15\), consistent with the primary simulation. The same prior specifications, posterior sampling settings, and implementations of EMP, BB, HBB, and AFT-BART-NP were used in both analyses.

\end{document}